\documentclass[iop]{emulateapj-rtx4}

\usepackage{mathrsfs }
\usepackage{natbib}
\usepackage{amssymb}
\usepackage{apjfonts}
\usepackage{ulem} %Makes striking out text possible with \sout{}
\begin{document}

\title{The MOSDEF Survey: Sulfur Emission-line Ratios Provide New Insights into Evolving ISM Conditions at High Redshift\altaffilmark{1}}

\author{
 Alice E. Shapley,\altaffilmark{2}
 Ryan L. Sanders,\altaffilmark{3}
 Peng Shao,\altaffilmark{4}
 Naveen A. Reddy,\altaffilmark{5}
 Mariska Kriek,\altaffilmark{6}	
 Alison L. Coil,\altaffilmark{7}
 Bahram Mobasher,\altaffilmark{5}	
 Brian Siana,\altaffilmark{5}
 Irene Shivaei,\altaffilmark{8,9}
 William R. Freeman,\altaffilmark{5}
 Mojegan Azadi,\altaffilmark{10}
 Sedona H. Price,\altaffilmark{11}
 Gene C. K. Leung,\altaffilmark{7}
 Tara Fetherolf,\altaffilmark{5}
 Laura de Groot,\altaffilmark{12}
 Tom Zick,\altaffilmark{6}
 Francesca M. Fornasini,\altaffilmark{10}
 Guillermo Barro\altaffilmark{13}
}

\altaffiltext{1}{Based on data obtained at the W.M. Keck Observatory, which is operated as a scientific partnership among the California Institute of Technology, the University of California,  and the National Aeronautics and Space Administration, and was made possible by the generous financial support of the W.M. Keck Foundation.}
\altaffiltext{2}{Department of Physics and Astronomy, University of California, Los Angeles, 430 Portola Plaza, Los Angeles, CA 90095, USA}
\altaffiltext{3}{Department of Physics, University of California, Davis, 1 Shields Avenue, Davis, CA 95616, USA}
\altaffiltext{4}{School of Astronomy and Space Science, Nanjing University, Nanjing 210023, People's Republic of China}
\altaffiltext{5}{Department of Physics and Astronomy, University of California, Riverside, 900 University Avenue, Riverside, CA 92521, USA}
\altaffiltext{6}{Astronomy Department, University of California at Berkeley, Berkeley, CA 94720, USA}
\altaffiltext{7}{Center for Astrophysics and Space Sciences, Department of Physics, University of California, San Diego, 9500 Gilman Drive., La Jolla, CA 92093, USA}
\altaffiltext{8}{Steward Observatory, University of Arizona, 933 N Cherry Ave, Tucson, AZ 85721, USA}
\altaffiltext{9}{Hubble Fellow}
\altaffiltext{10}{Harvard-Smithsonian Center for Astrophysics, 60 Garden Street, Cambridge, MA, 02138, USA}
\altaffiltext{11}{Max-Planck-Institut f\"ur Extraterrestrische Physik, Postfach 1312, Garching, 85741, Germany}
\altaffiltext{12}{Department of Physics, The College of Wooster, 1189 Beall Avenue, Wooster, OH 44691, USA}
\altaffiltext{13}{Department of Phyics, University of the Pacific, 3601 Pacific Ave, Stockton, CA 95211, USA}
\email{aes@astro.ucla.edu}

\shortauthors{Shapley et al.}

% Title for running header

\shorttitle{Sulfur and the ISM at High Redshift}

%****************************************************************** %

\begin{abstract} 
We present results on the emission-line properties 
of $1.3\leq z \leq 2.7$ galaxies drawn from the complete MOSFIRE
Deep Evolution Field (MOSDEF) survey. Specifically, we use observations
of the emission-line diagnostic diagram of [OIII]$\lambda 5007$/H$\beta$ vs.
[SII]$\lambda\lambda6717,6731$/H$\alpha$, i.e., the ``[SII] BPT diagram," to gain insight
into the physical properties of high-redshift star-forming regions. 
High-redshift MOSDEF galaxies are offset towards lower
[SII]$\lambda\lambda6717,6731$/H$\alpha$ at fixed [OIII]$\lambda 5007$/H$\beta$, relative
to local galaxies from the Sloan Digital Sky Survey (SDSS). Furthermore, at fixed [OIII]$\lambda 5007$/H$\beta$,
local SDSS galaxies follow a trend of decreasing [SII]$\lambda\lambda6717,6731$/H$\alpha$
as the surface density of star formation ($\Sigma_{\rm{SFR}}$)
increases. We explain this trend in terms of the decreasing fractional contribution
from diffuse ionized gas ($f_{\rm{DIG}}$) as $\Sigma_{\rm{SFR}}$ increases in galaxies, which causes galaxy-integrated
line ratios to shift towards the locus of pure H~II-region emission. The $z\sim0$ relationship between
$f_{\rm{DIG}}$ and $\Sigma_{\rm{SFR}}$ implies that high-redshift galaxies have lower
$f_{\rm{DIG}}$ values than typical local systems, given their significantly higher typical $\Sigma_{\rm{SFR}}$.
When an appropriate low-redshift benchmark with zero or minimal $f_{\rm{DIG}}$ is used, high-redshift MOSDEF galaxies appear offset
towards higher [SII]$\lambda\lambda6717,6731$/H$\alpha$ and/or [OIII]$\lambda 5007$/H$\beta$.
The joint shifts of high-redshift galaxies in the [SII] and [NII] BPT diagrams are best
explained in terms of the harder spectra ionizing their star-forming regions at fixed nebular oxygen abundance 
(expected for chemically-young galaxies), as opposed to large variations in N/O ratios or higher ionization
parameters. The evolving mixture of H~II regions and DIG is an essential ingredient to our description
of the ISM over cosmic time.
\end{abstract} 

%****************************************************************** %


\section{Introduction}
\label{sec:introduction}
Rest-optical emission line ratios provide a powerful probe
of the physical conditions in the interstellar medium (ISM).
Local star-forming galaxies trace a tight sequence of increasing
[NII]$\lambda6584$/H$\alpha$ and decreasing [OIII]$\lambda5007$/H$\beta$,
as metallicity increases and the overall excitation in star-forming regions
decreases. Recent statistical samples of rest-optical spectra of $z\sim2$
galaxies show that high-redshift galaxies
are offset systematically from local galaxies towards higher [OIII]$\lambda5007$/H$\beta$ and
[NII]$\lambda6584$/H$\alpha$ values on average \citep{steidel2014,shapley2015}. There are many possible causes for this observed
difference in $z\sim2$ galaxies, including higher ionization
parameters in distant H~II regions, harder ionizing spectra at fixed metallicity for the stars
photo-ionizing the H~II regions, higher densities (or equivalently pressures), variations in the gas-phase N/O 
abundance patterns, and contributions from active galactic nuclei (AGNs) and shocks
\citep[e.g.,][]{kashino2017,steidel2016,sanders2016,masters2014,coil2015,freeman2019}.

The differences between high-redshift and local emission-line sequences were first
noted in the [OIII]$\lambda5007$/H$\beta$  vs. [NII]$\lambda6584$/H$\alpha$ diagnostic
diagram, i.e., the so-called [NII] BPT diagram \citep{baldwin1981}. However,
in interpreting these differences, various authors have considered
the properties of galaxies in the [OIII]$\lambda5007$/H$\beta$ vs. [SII]$\lambda\lambda 6717,6731$/H$\alpha$
diagram (first introduced in \citet{veilleux1987}, and referred to hereafter as the ``[SII] BPT diagram")
and the space of [OIII]$\lambda\lambda$4959,5007/[OII]$\lambda\lambda$3726,3729 ($O_{32}$)
vs. ([OIII]$\lambda\lambda$4959,5007+[OII]$\lambda\lambda$3726,3729)/H$\beta$ ($R_{23}$).
The lack of a significant positive offset in the [SII] BPT and $O_{32}$ vs. $R_{23}$ diagrams
has been used to argue that $z\sim0$ and $z\sim2$ galaxies follow different
N/O abundance patterns \citep{masters2014,shapley2015}. A small positive offset measured in
the [SII] BPT diagram has been used to suggest a harder ionizing spectrum at fixed
nebular abundance \citep{strom2017,steidel2016} at $z\sim2$. Meanwhile, a {\it negative} offset
in [SII]$\lambda\lambda6717,6731$/H$\alpha$ at fixed
[OIII]$\lambda5007$/H$\beta$ for $z\sim1.5$ star-forming galaxies has been attributed
to a higher ionization parameter \citep{kashino2017,kashino2019}.

While arriving at different conclusions regarding the evolution of the properties
of star-forming regions at high redshift, the analyses of local and high-redshift
emission-line diagnostic diagrams typically share a common approach. Specifically,
integrated slit or fiber spectra of distant galaxies are compared with 
fiber spectra of local galaxies drawn from the Sloan Digital Sky Survey \citep[SDSS;][]{abazajian2009}.
Each galaxy is effectively represented as a point source, when in fact the integrated
spectrum of a galaxy contains the sum of the emission from the ensemble of H~II regions
that fall within the spectral aperture, along with the contribution from diffuse ionized
gas (DIG) in the ISM. DIG exists outside H~II regions, and has been shown to
contribute typically 30\%-60\%  of the total H$\alpha$ flux in local
spiral galaxies \citep{zurita2000,oey2007}. Furthermore, the fractional contribution of
DIG emission to the Balmer lines declines with increasing star-formation rate (SFR)
surface density \citep[$\Sigma_{\rm{SFR}}$;][]{oey2007}. \citet{zhang2017} and \citet{sanders2017} have
also demonstrated that distinct physical conditions and ionizing spectra in the DIG and H~II regions
result in systematically different DIG and H~II region excitation sequences in emission-line diagrams
featuring [SII] or [OII], while not strongly affecting the [NII] BPT diagram.
Clearly, a robust interpretation of the evolving internal properties of H~II regions with increasing redshift
requires an understanding of the evolving mixture of DIG and H~II regions within star-forming
galaxies.

In this work, we examine how the different mixtures of DIG and H~II
region emission at low and high redshifts affect inferences regarding the evolution of
H~II region properties out to $z\sim2$.  We analyze galaxies at $1.3\leq z\leq2.7$ drawn
from the MOSFIRE Deep Evolution Field (MOSDEF) survey \citep{kriek2015} in comparison with
local SDSS galaxies.  Most importantly, we consider the implications of the fact that $\Sigma_{\rm{SFR}}$
is typically two orders of magnitude higher
in star-forming galaxies at $z\sim2$, compared with $z\sim0$ galaxies.
In \S\ref{sec:observations}, we describe
our observations and samples. In \S\ref{sec:emline}, we  revisit the comparison
between local and high-redshift emission-line diagrams, accounting for their additional
differences in $\Sigma_{\rm{SFR}}$. In \S\ref{sec:discussion}, we discuss
the implications for inferring the internal properties of high-redshift star-forming
regions.  Throughout, we adopt cosmological parameters of 
$H_0=70\mbox{ km  s}^{-1}\mbox{ Mpc}^{-1}$, $\Omega_m=0.30$, and
$\Omega_{\Lambda}=0.7$, and a \citet{chabrier2003} IMF.

\begin{figure*}
\centering
\includegraphics[width=0.95\textwidth]{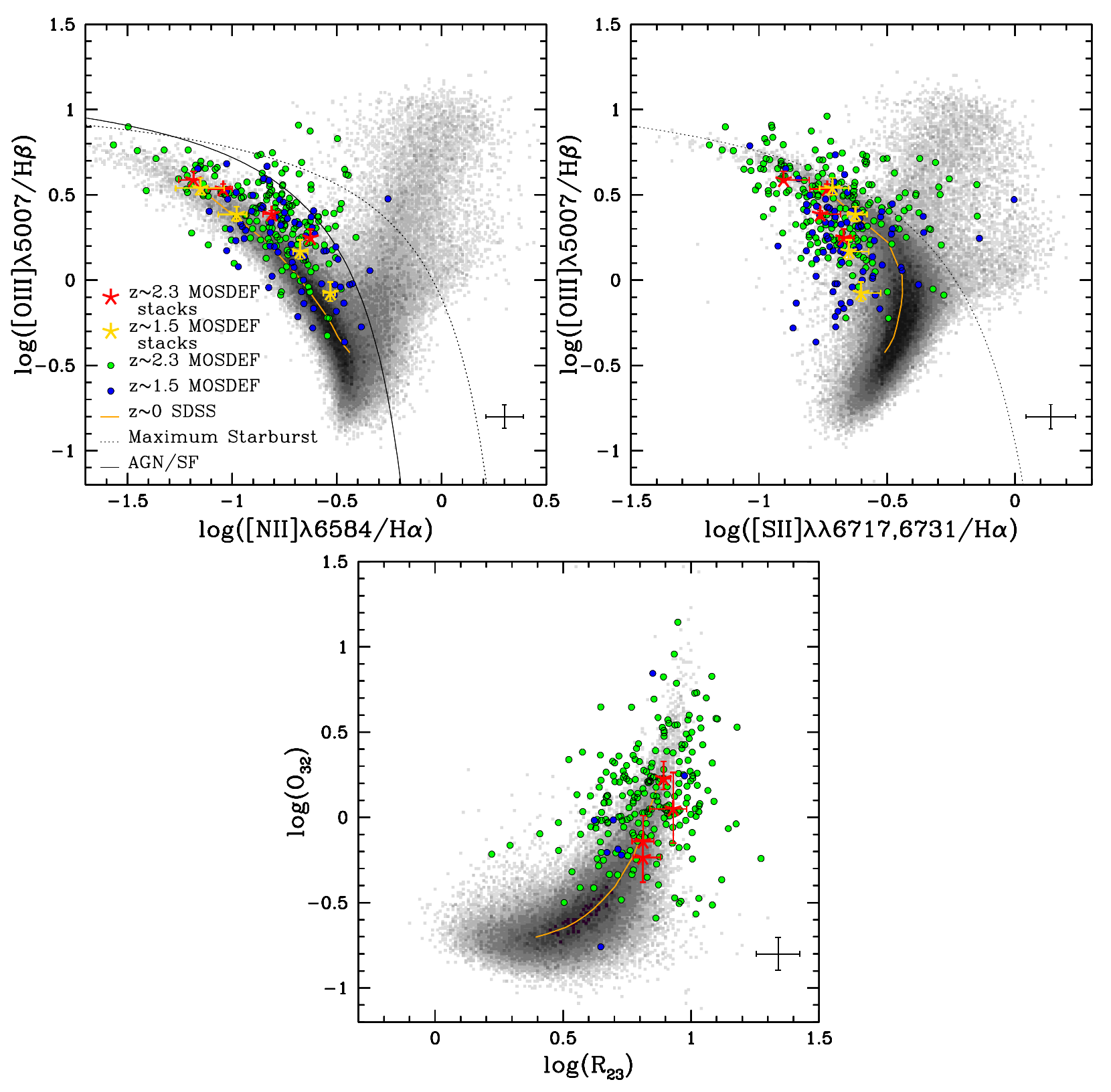}
\caption{{\bf Top left:} [NII] BPT diagram for $1.4\leq z\leq2.7$ MOSDEF galaxies. Green [blue] points indicate
$z\sim2.3$ [$z\sim1.5$] MOSDEF galaxies with $\geq3\sigma$ detections of all 4 BPT emission lines. Median
MOSDEF errorbars are indicated in the lower-right-hand corner of each panel.
The grayscale histogram and orange curve correspond, respectively, to the distribution and running median of local SDSS galaxies.
The running median line ratios are calculated in closely-spaced bins of stellar mass.
Large red [gold] stars indicate measurements from composite spectra, binned by stellar mass, of all MOSDEF
$z\sim2.3$ [$z\sim1.5$] galaxies with coverage of the relevant emission lines and $\geq3\sigma$ H$\alpha$ detections.
Stacks of increasing stellar mass have lower [OIII]$\lambda 5007$/H$\beta$. The black dotted
curve is the ``maximum starburst" line from \citet{kewley2001}, while the black solid curve is an empirical
AGN/star-formation threshold from \citet{kauffmann2003}. Although plotted here for completeness,
SDSS galaxies falling above the \citet{kauffmann2003} curve are not included in our analysis.
{\bf Top right:} [SII] BPT diagram. Symbols are the same
as in the left-hand panel. {\bf Bottom:} $O_{32}$ vs. $R_{23}$ diagram, corrected for dust.  Stacked points are not 
shown for the $z\sim1.5$ sample in this panel, given the small size of the sample with [OII]$\lambda\lambda3726,3729$
coverage.
}
\label{fig:bptpanels}
\end{figure*}

\section{Observations and Samples}
\label{sec:observations}

\subsection{MOSDEF Survey and Sample}
\label{sec:observations-mosdef}

Our analysis is based on the complete MOSDEF survey data set.
Full details of the survey observations and data reduction
are provided in \citet{kriek2015}. In brief, MOSDEF was a 48.5-night observing program
over 4 years using the Multi-Object Spectrometer for Infra-Red
Exploration \citep[MOSFIRE;][]{mclean2012} on the 10~m
Keck~I telescope. With MOSDEF, we obtained rest-optical
spectra for a sample of $\sim1500$ galaxies at $1.4\leq z\leq3.8$
in the COSMOS, GOODS-N, AEGIS, GOODS-S, and UDS fields covered by the 
CANDELS and 3D-HST surveys \citep{grogin2011,koekemoer2011,momcheva2016}. These fields 
are covered by extensive multi-wavelength observations
\citep[e.g.,][]{skelton2014}. MOSDEF targets fall within three
distinct redshift intervals, where the strongest rest-optical emission
lines are accessible through windows of atmospheric transmission:
$1.37\leq z\leq 1.70$, $2.09\leq  z\leq 2.61$, and $2.95\leq z\leq 3.80$.
Here we focus on galaxies in the two lower-redshift intervals, hereafter
described as $z\sim1.5$ and $z\sim2.3$, for which not only 
[OII]$\lambda\lambda3726,3729$, H$\beta$, [OIII]$\lambda\lambda 4959,5007$
are accessible from the ground, but also H$\alpha$,
[NII]$\lambda6584$, and [SII]$\lambda\lambda6717,6731$.

In addition to measurements of strong rest-optical nebular
emission lines for MOSDEF galaxies, 
we also analyze dust-corrected 
H$\alpha$ star-formation surface densities, i.e., $\Sigma_{\rm{SFR}}$.
To obtain $\Sigma_{\rm{SFR}}$, we determined nebular extinction,
$E(B-V)_{{\rm neb}}$, from the stellar-absorption-corrected H$\alpha$/H$\beta$ Balmer decrement and the 
assumption of the Milky Way dust extinction curve \citep{cardelli1989}. H$\alpha$ SFRs  (SFR(H$\alpha$))
were then estimated from dust-corrected and slit-loss-corrected H$\alpha$ luminosities, based on the
calibration of \citet{hao2011} for a \citet{chabrier2003} IMF. The procedures for stellar-absorption, dust,
and slit-loss corrections are fully described in \citet{reddy2015} and \citet{kriek2015}. We also require
galaxy sizes for $\Sigma_{\rm{SFR}}$. \citet{vanderwel2014} fit single-component S\'ersic profiles to the two-dimensional
light distributions of galaxies in the CANDELS fields, and derive half-light radii, $r_{\rm{e}}$,
as the semi-major axis of the ellipse containing half of the total galaxy light. We use F160W galaxy
half-light radii from the publicly-available catalogs of \citeauthor{vanderwel2014}, and then define SFR surface
density as:

\begin{equation}
\Sigma_{\rm{SFR}} = \frac{\mbox{SFR(H}\alpha\mbox{)}}{2\pi r_{\rm{e}}^2}
\label{eq:sigmasfr}
\end{equation}

Our main analysis is based on samples of MOSDEF galaxies at $z\sim1.5$
and $z\sim2.3$ with coverage of the H$\beta$, [OIII], H$\alpha$, [NII], and [SII]
emission lines. Specifically, we selected 434 [211] galaxies at $2.0\leq z\leq2.7$
[$1.3\leq z\leq1.7$], with robust spectroscopic redshifts, S/N~$\geq3$
in H$\alpha$ emission-line flux, and no evidence for AGN activity
based on X-ray luminosity, {\it Spitzer}/IRAC colors, or [NII]$\lambda6584$/H$\alpha$
ratios \citep{coil2015,azadi2017}. The $z\sim2.3$ sample is characterized by
a median redshift and stellar mass of $z_{\rm{med}}=2.28$ and $\log(M/M_{\odot})_{\rm{med}}=9.92$, respectively,
while the corresponding values for the $z\sim1.5$ sample are
$z_{\rm{med}}=1.52$ and $\log(M/M_{\odot})_{\rm{med}}=9.95$. 

\subsection{SDSS $z\sim0$ Comparison Sample}
\label{sec:observations-sdss}
In order to gain insights into the evolving properties
of star-forming regions at high redshift, we selected
a comparison sample of local galaxies from the Sloan
Digital Sky Survey (SDSS) Data release 7
\citep[DR7;][]{abazajian2009}. Stellar-absorption-corrected emission-line measurements
are drawn from the MPA-JHU catalog of measurements
for DR7\footnote{Available at http://www.mpa-garching.mpg.de/SDSS/DR7/},
as well as the corrections needed to correct fiber to total emission-line fluxes.
In order to estimate $\Sigma_{\rm{SFR}}$ for SDSS galaxies,
we calculated $E(B-V)_{{\rm neb}}$, dust-corrected H$\alpha$ luminosities, and
SFR(H$\alpha$) for SDSS galaxies using the methodology described above.
In analogy with the rest-optical half-light radii adopted for MOSDEF galaxies,
we used the elliptical Petrosian $R$-band half-light radii for SDSS galaxies
drawn from the NASA-Sloan Atlas v1.0.1\footnote{Available at http://www.nsatlas.org.},
and applied equation~\ref{eq:sigmasfr} to obtain $\Sigma_{\rm{SFR}}$. We restricted the SDSS
sample to galaxies at $0.04\leq z\leq0.10$ to reduce aperture effects, and 
required 5$\sigma$ detections for all lines included
in the [SII] and [NII] BPT diagrams (i.e., H$\beta$, [OIII], H$\alpha$, [NII], and [SII]).
As we wish to study the star-formation properties of the SDSS sample, we also
removed galaxies from our analysis that satisfied the optical emission-line AGN criterion of 
\citet{kauffmann2003}. We finally required a robust half-light radius entry in the 
NASA-Sloan Atlas. The above criteria yielded a $z\sim0$ comparison sample of 60,609 SDSS galaxies 
with $\log(M/M_{\odot})_{\rm{med}}=9.83$.

\section{Emission-Line Ratios at Low and High Redshift}
\label{sec:emline}

\begin{figure}[t!]
\includegraphics[width=0.5\textwidth]{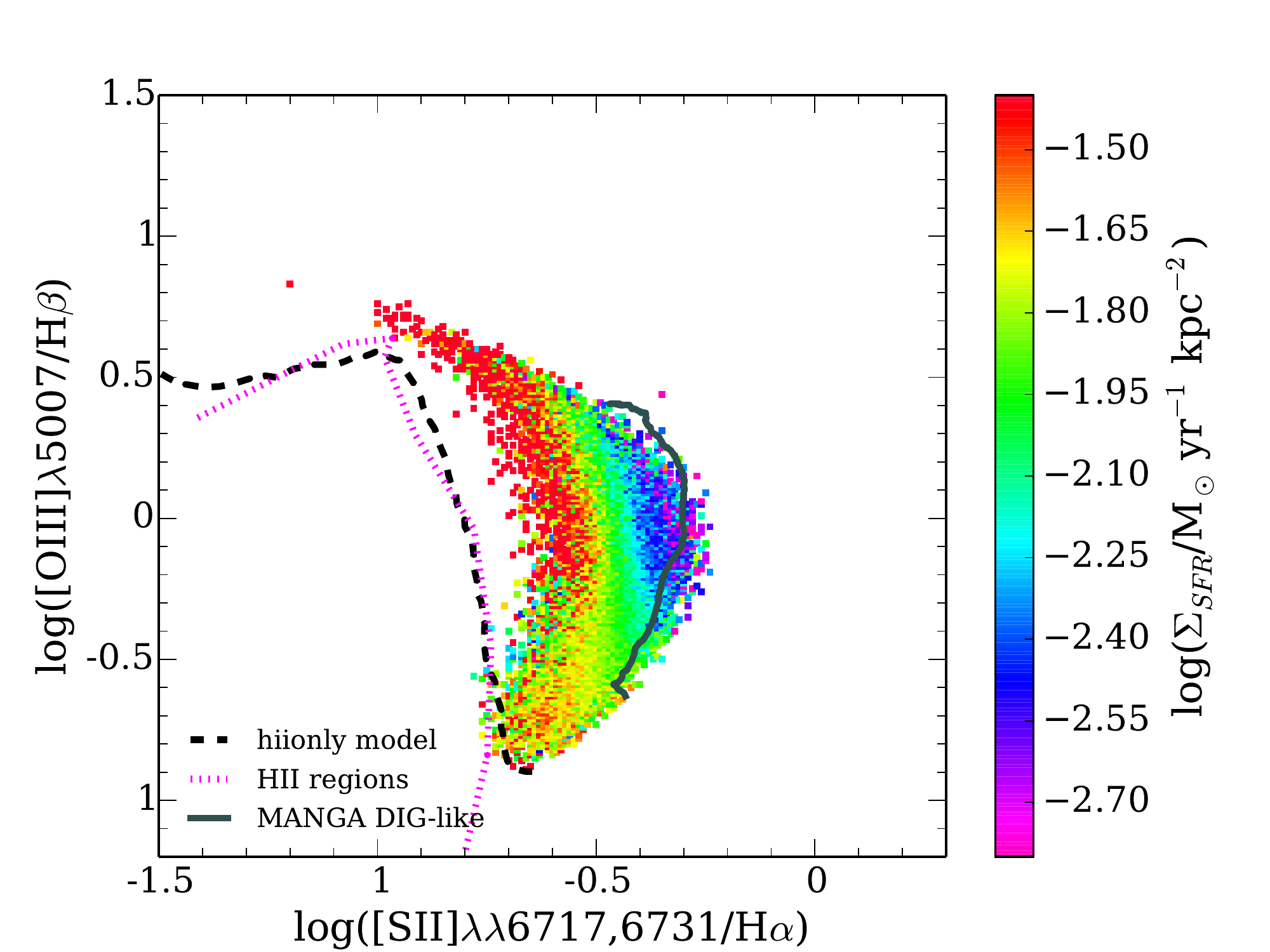}
\caption{Median $\Sigma_{\rm{SFR}}$ for SDSS galaxies
as a function of position in the [SII] BPT diagram. There is a strong trend for galaxies with higher values
of $\Sigma_{\rm{SFR}}$ to be shifted towards lower values of [SII]$\lambda\lambda6717,3731$/H$\alpha$ at fixed
[OIII]$\lambda 5007$/H$\beta$ \citep{masters2016}. In addition, we indicate the running median for local
H~II regions \citep{pilyugin2016} with the dotted pink line. The running median line ratios are calculated in bins of 
H~II-region electron temperature.  The {\it hiionly} model from \citet{sanders2017}
is shown with the dashed black line, representing the ensemble average emission from H~II regions
in star-forming galaxies in the absence of DIG emission. Finally, the running median for ``DIG"-like
(i.e., low H$\alpha$ surface brightness) spaxels from the SDSS/MaNGA DIG galaxy sample used in \citet{sanders2017}
is shown as the solid dark-grey curve. The running median line ratios for the ``DIG"-like curve are calculated in bins of 
([OIII]$\lambda 5007$/H$\beta$)/([NII]$\lambda6584$/H$\alpha$), which correlates with nebular metallicity in local H~II regions
\citep{pettinipagel2004}.
}
\label{fig:o3s2-ss}
\end{figure}

\begin{figure*}[t!]
\centering
\includegraphics[width=0.6\textwidth]{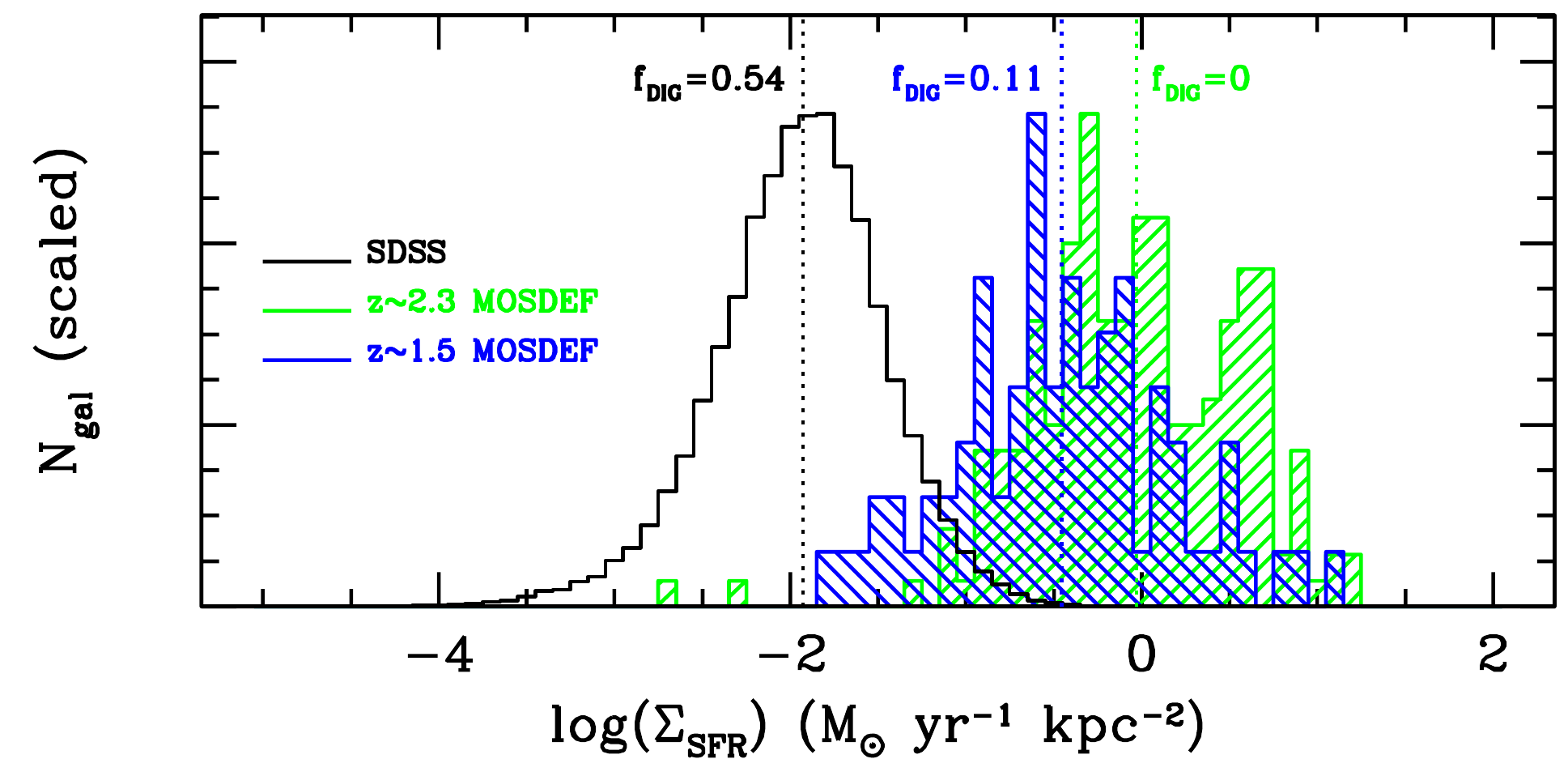}
\caption{$\Sigma_{\rm{SFR}}$ distributions for SDSS and MOSDEF galaxies. The black histogram indicates
the distribution in $\Sigma_{\rm{SFR}}$ for the $z\sim0$ SDSS sample, while the shaded green [blue]
histogram indicates the corresponding distribution for $z\sim2.3$ [$z\sim1.5$] MOSDEF galaxies.
Equation~\ref{eq:fdig} is used to translate the median $\Sigma_{\rm{SFR}}$ of each distribution to a corresponding
$f_{\rm{DIG}}$. While local star-forming galaxies typically have $f_{\rm{DIG}}=0.54$, the median
$\Sigma_{\rm{SFR}}$ of $z\sim2.3$ MOSDEF galaxies, which is almost two orders of magnitude higher,
suggests $f_{\rm{DIG}}=0$.
}
\label{fig:sigmasfr-hist}
\end{figure*}

\subsection{MOSDEF Emission-Line Diagrams}
\label{sec:emline-mosdef}
We present here a significantly expanded sample of $z\sim2.3$ rest-optical emission-line
ratio measurements compared to our previous work \citep{shapley2015,sanders2016}, now
based on the completed MOSDEF survey. In addition, for the first
time we present the same set of measurements for the $z\sim1.5$ MOSDEF sample.
Figure~\ref{fig:bptpanels} shows MOSDEF galaxies in the [NII] and [SII] BPT
diagrams, as well as $O_{32}$ vs. $R_{23}$. Of the
galaxies in the [NII] and [SII] BPT samples with coverage of all BPT features
and detections of H$\alpha$ emission, there are 179 [68] objects with all
[NII] BPT features individually detected at $\geq3\sigma$ significance
at $z\sim2.3$ [$z\sim1.5$]. The $z\sim2.3$ [$z\sim1.5$] sample size with all features
detected in the [SII] BPT diagram is 201 [79]. For the $O_{32}$ vs. $R_{23}$ diagram
the number of galaxies with [OII], H$\beta$, [OIII], and H$\alpha$ detected is 224 [8]
at $z\sim2.3$ [$z\sim1.5$]. In each of the 3 panels of Figure~\ref{fig:bptpanels},
we show not only the subset of individual detections, but measurements from median stacked spectra from the full sample
of MOSDEF galaxies in bins of stellar mass, constructed as described in \citet{sanders2018}.

Based on the full MOSDEF sample, we recover the well-known offset for $z>1$ galaxies
in the [NII] BPT diagram, towards higher [NII]$\lambda6584$/H$\alpha$ and [OIII]$\lambda5007$/H$\beta$
relative to the sequence of $z\sim0$ star-forming galaxies. In detail, the $z\sim2.3$ MOSDEF sample
appears to be slightly more offset than the $z\sim1.5$ sample. In the [SII] BPT diagram,
the spectral stacks reveal the average trend for $z\geq1.3$ MOSDEF galaxies is offset from the local sequence
towards {\it lower} [SII]$\lambda\lambda6717,6731$/H$\alpha$ at fixed [OIII]$\lambda5007$/H$\beta$,
consistent with the results of \citet{kashino2017,kashino2019}. In the $O_{32}$ vs. $R_{23}$ diagram,
we highlight the results from $z\sim2.3$, since the $z\sim1.5$ sample contains few
objects with coverage of [OII]$\lambda\lambda3726,3729$. Spectral
stacks show that the $z\sim2.3$ sequence is slightly offset towards higher $R_{23}$ at fixed
$O_{32}$, relative to the local sequence. In the discussion that follows, we focus on the [SII]
BPT diagram in more detail.

\subsection{$\Sigma_{\rm{SFR}}$ and the [SII] BPT Diagram}
\label{sec:emline-siisigma}
In order to understand the emission-line properties of high-redshift galaxies, 
we turn to the local [SII] BPT diagram, considering the connection between the location in this diagram
and $\Sigma_{\rm{SFR}}$. As previously featured in \citet{masters2016}, Figure~\ref{fig:o3s2-ss} shows
a clear trend towards lower [SII]$\lambda\lambda6717,6731$/H$\alpha$ at fixed [OIII]$\lambda5007$/H$\beta$
as  $\Sigma_{\rm{SFR}}$ increases. We now explain this trend based on the result from \cite{oey2007},
according to which the fraction of the Balmer flux contributed by DIG, $f_{\rm{DIG}}$, is a decreasing
function of the dust-corrected H$\alpha$ surface brightness -- or, equivalently, $\Sigma_{\rm{SFR}}$.
Recasting the fit from \citet{sanders2017} to the \citeauthor{oey2007} data in terms of $\Sigma_{\rm{SFR}}$,
we find:

\begin{equation}
f_{\rm{DIG}}=-0.900 \times (\frac{\Sigma_{\rm{SFR}}}{M_{\odot}\mbox{yr}^{-1}\mbox{kpc}^{-2}})^{1/3}+0.748
\label{eq:fdig}
\end{equation}

Accordingly, galaxies with higher $\Sigma_{\rm{SFR}}$ have lower $f_{\rm{DIG}}$. In addition
to showing the [OIII]$\lambda5007$/H$\beta$ vs. [SII]$\lambda\lambda6717,6731$/H$\alpha$
locus for SDSS galaxies, Figure~\ref{fig:o3s2-ss} 
also features the corresponding median sequence for local H~II regions drawn from the sample of \citet{pilyugin2016},
and that traced out by the {\it hiionly} model from \citet{sanders2017}. The {\it hiionly} model represents
the ensemble average emission from H~II regions in star-forming galaxies, and does not include DIG emission.
Figure~\ref{fig:o3s2-ss} also shows the median sequence for DIG-dominated spaxels from galaxies in the SDSS MaNGA
survey \citep{zhang2017}. The H~II region sequences are offset towards significantly {\it lower} [SII]$\lambda\lambda6717,6731$/H$\alpha$
at fixed [OIII]$\lambda5007$/H$\beta$, relative to SDSS galaxies, while the DIG-dominated spaxels are shifted towards
{\it higher} [SII]$\lambda\lambda6717,6731$/H$\alpha$. It is worth noting that the same segregation between
H~II region, SDSS galaxy, and DIG emission sequences is not apparent in the [NII] BPT diagram \citep{zhang2017,sanders2017}.
It is clear that, as $\Sigma_{\rm{SFR}}$ increases and $f_{\rm{DIG}}$ decreases, the 
[OIII]$\lambda5007$/H$\beta$ vs. [SII]$\lambda\lambda6717,6731$/H$\alpha$ locus of SDSS galaxies shifts towards
the H~II region sequence (or, equivalently, the {\it hiionly} model, which excludes the contribution of DIG emission).

\subsection{$\Sigma_{\rm{SFR}}$ at Low and High Redshift}
\label{sec:emline-evolsigma}
At $z>1$, star-forming galaxies of the same stellar mass
have significantly higher SFRs and smaller radii. The combination
of these factors results in dramatically different median values of $\Sigma_{\rm{SFR}}$
for the MOSDEF samples presented here, compared with the properties of $z\sim0$ SDSS galaxies.
In Figure~\ref{fig:sigmasfr-hist}, we show the distribution of $\Sigma_{\rm{SFR}}$
for our $z\sim0$ SDSS comparison sample (black histogram), as well as the corresponding
distributions for the $z\sim2.3$ and $z\sim1.5$ MOSDEF samples (green and blue shaded
histograms, respectively). The median $\log(\Sigma_{\rm{SFR}})$ values at $z\sim0$, 1.5 and 2.3
are $-1.93$, $-0.46$, and $-0.03$, where $\Sigma_{\rm{SFR}}$ is in units of $M_{\odot}\mbox{ yr}^{-1}\mbox{ kpc}^{-2}$.
Therefore, the difference in typical $\Sigma_{\rm{SFR}}$ at $z\sim0$ and $z\sim2.3$ is almost two orders
of magnitude. We have also used equation~\ref{eq:fdig} to estimate the corresponding $f_{\rm{DIG}}$
value for each median $\Sigma_{\rm{SFR}}$, finding median $f_{\rm{DIG}}$ values of 0.54, 0.11, and 0, for
at $z\sim0$, 1.5, and 2.3. For these calculations, we are making the simplifying assumption
that the same relation between $f_{\rm{DIG}}$ and $\Sigma_{\rm{SFR}}$ holds at low and high redshift, although
it has only been measured locally. If this assumption is valid, then the emission from the ionized
ISM at $z\geq 1.5$ should be well represented by the ensemble-averaged emission of H~II regions,
with minimal (or zero, in the case of $z\sim2.3$) contribution from DIG. This difference in
the relative contributions of H~II regions and DIG must be accounted for when interpreting
the [SII] BPT diagram of high-redshift galaxies.

\subsection{Revisiting the High-Redshift [SII] BPT Diagram}
\label{sec:emline-revisit}
\begin{figure*}[t!]
\centering
\includegraphics[width=0.95\textwidth]{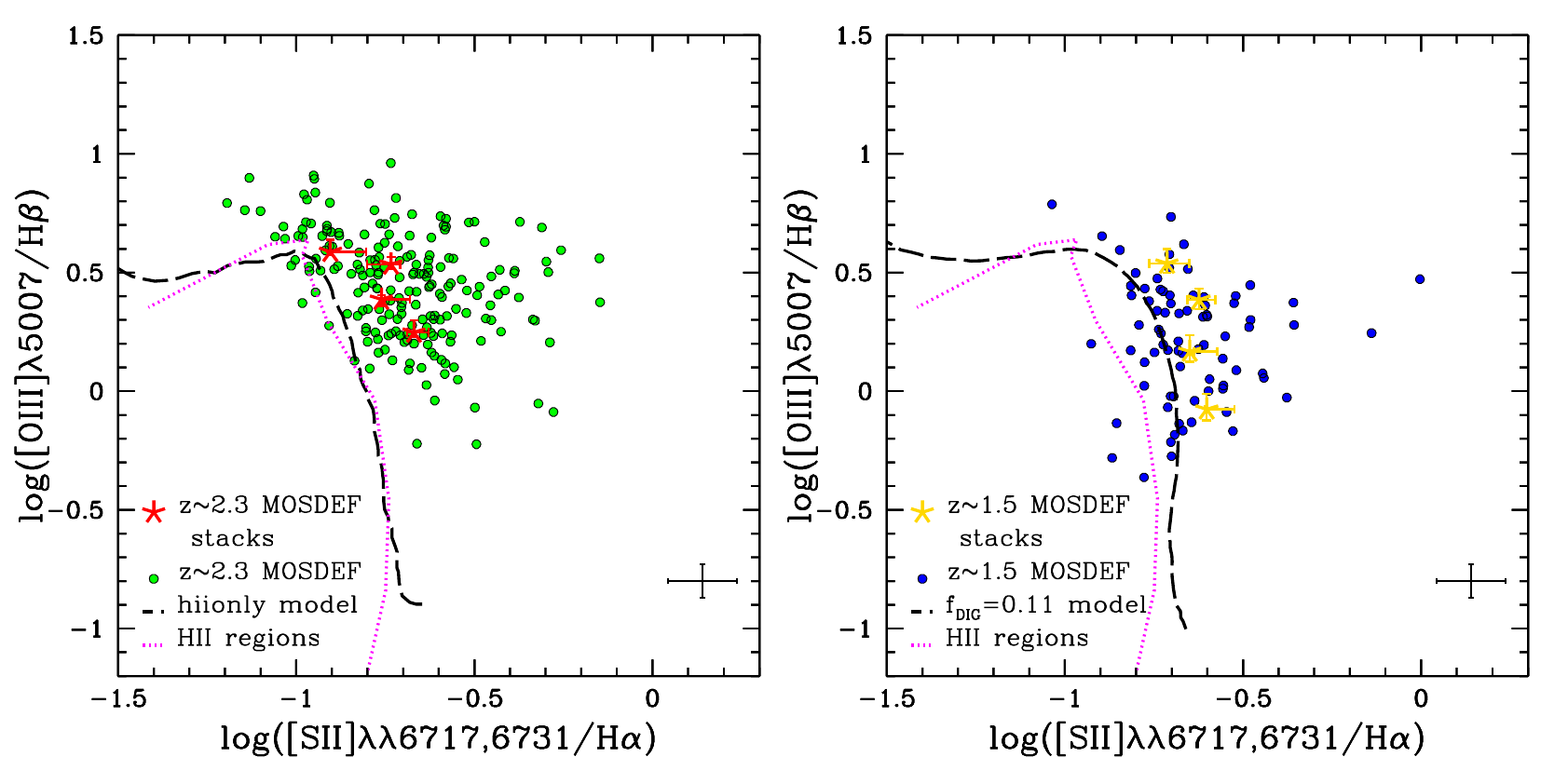}
\caption{Comparison in the [SII] BPT diagram beween MOSDEF galaxies, HII~regions, and models with low DIG emission fraction.
{\bf Left:} [SII] BPT diagram for $z\sim2.3$ MOSDEF galaxies and stacks (green points, red stars), median sequence of local H~II regions (magenta dotted
line), and {\it hiionly} model from \citet{sanders2017} (black dashed line). $z\sim2.3$ MOSDEF galaxies are clearly offset towards
larger [OIII]$\lambda 5007$/H$\beta$ and/or larger [SII]$\lambda\lambda6717,6731$/H$\alpha$ relative to local H~II regions
and the {\it hiionly} model. {\bf Right:} The same plot, but for $z\sim1.5$ MOSDEF galaxies and stacks (blue points, gold stars), local H~II regions
(magenta dotted line), and a model from \citet{sanders2017} representing the ensemble average emission from H~II regions
in star-forming galaxies plus a DIG fractional contribution to the H$\alpha$ emission of $f_{\rm{DIG}}=0.11$. The offset
between $z\sim1.5$ MOSDEF galaxies and the comparison curves is smaller than for the $z\sim2.3$ MOSDEF sample.
}
\label{fig:hii-mosdef}
\end{figure*}

In order to infer the differences between low and high-redshift star-forming regions, we
need to consider low- and high-redshift samples with the same $f_{\rm{DIG}}$. For $z\sim2.3$
MOSDEF galaxies, the appropriate comparison sample is either local H~II regions, or the 
{\it hiionly} model of \citet{sanders2017}, with $f_{\rm{DIG}}=0$. Given the slightly lower $\Sigma_{\rm{SFR}}$ and higher
inferred $f_{\rm{DIG}}$ for the MOSDEF $z\sim1.5$ sample, we generated a model with $f_{\rm{DIG}}=0.11$ for the purposes
of comparison, based on the methodology of \citet{sanders2017}. Figure~\ref{fig:hii-mosdef} shows
the [SII] BPT diagrams for the $z\sim2.3$ and $z\sim1.5$ MOSDEF samples of individual detections,
along with the appropriate comparison models. The high-redshift MOSDEF samples are clearly shifted
towards higher [SII]$\lambda\lambda6717,6731$/H$\alpha$ and [OIII]$\lambda5007$/H$\beta$ relative
to the low-DIG comparison models, with a larger positive shift for the $z\sim2.3$ sample.

\section{Discussion}
\label{sec:discussion}

Previous comparisons between high-redshift and local SDSS galaxies in the [SII] BPT diagram
suggested that there was zero or {\it negative} shift in [SII]$\lambda\lambda6717,6731$/H$\alpha$
at fixed [OIII]$\lambda5007$/H$\beta$. Such results led \citet{shapley2015}
to conclude that differences existed between low- and high-redshift N/O vs. O/H abundance patterns,
and \citet{kashino2017} to infer higher ionization parameters at fixed nebular abundance.
However, such analyses did not use the appropriate $z\sim 0$ comparison sample with minimal $f_{\rm{DIG}}$.
Our revised comparison in the [SII] BPT diagram between 
high-redshift galaxies and local H~II regions, or galaxies with
minimal DIG contribution, demonstrates definitively
that $z\geq1.3$ galaxies have significant positive x- and/or y-offsets relative
to their appropriate local counterparts in {\it both} the [NII] and [SII] BPT diagrams.
\citet{sanders2016} showed that, compared
to a fiducial set of photoionization models, those with harder ionizing spectrum
at fixed metallicity lead to corresponding positive offsets in both [NII] and [SII] BPT diagrams.
\citet{steidel2016} proposed an underlying physical scenario for such behavior, according to which chemically-young
high-redshift galaxies exhibit $\alpha$ enhancement, with stellar Fe-based metallicities a factor of several lower than
nebular oxygen-based metallicities. 
Our analysis shows that the [NII] and [SII] BPT diagrams of high-redshift galaxies can be jointly explained by the
combination of a harder ionizing spectrum at fixed metallicity due to $\alpha$ enhancement, and a lower $f_{\rm{DIG}}$, 
compared with local star-forming galaxies.

The evolving mixture of DIG and H~II region emission has implications for both
photoionization models and strong-line metallicity calibrations.
For example, the photoionization models of \citet{levesque2010} and \citet{hirschmann2017} 
neglect a detailed treatment of the DIG contribution to integrated galaxy line ratios, and neither set of 
model grids overlaps the bulk of $z\sim0$ galaxies that they aim to describe in the [SII] BPT diagram. 
In both cases, the model grids fall much closer
to the locus of H~II regions, and clearly require the addition of DIG emission to correctly
describe local galaxies \citep{sanders2017}. As for metallicity calibrations,
the recently-introduced ``N2S2" indicator relates metallicity to a linear combination of $\log(\mbox{[NII]/[SII]})$
and $\log(\mbox{[NII]/H}\alpha)$ \citep{dopita2016}. However, because it includes [SII], N2S2 is sensitive
to {\it both} metallicity and $f_{\rm{DIG}}$.
Robust evolutionary comparisons between
the integrated oxygen abundances of high- and low-redshift galaxies must be made using empirical indicators
that are insensitive to variations in  $f_{\rm{DIG}}$.

Our conclusions are based on the assumption that high-redshift galaxies follow the same
relationship between $f_{\rm{DIG}}$ and $\Sigma_{\rm{SFR}}$ that is observed in local galaxies \citep{oey2007}.
We now need to test our assumption of redshift invariance by constructing rest-optical emission-line 
maps of distant galaxies on sub-kpc scales, and measuring both $f_{\rm{DIG}}$ and $\Sigma_{\rm{SFR}}$. 
Such observations will be possible with the integral-field
unit (IFU) of the NIRSpec instrument aboard the {\it James Webb Space Telescope} and planned adaptive-optics-assisted
IFUs on the next generation of extremely large ground-based telescopes.

\section*{Acknowledgements}
We acknowledge support from NSF AAG grants AST-1312780, 1312547, 1312764, and 1313171, grant AR-13907
from the Space Telescope Science Institute, and grant NNX16AF54G from the NASA ADAP program. 
We also acknowledge a NASA contract supporting the ``WFIRST Extragalactic Potential Observations (EXPO) 
Science Investigation Team" (15-WFIRST15-0004), administered by GSFC.
We thank the 3D-HST collaboration, who provided us with spectroscopic
and photometric catalogs used to select MOSDEF targets and derive
stellar population parameters. 
We acknowledge the First Carnegie Symposium
in Honor of Leonard Searle for useful information and discussions
that benefited this work.  
We finally wish to extend special thanks to those of Hawaiian ancestry on
whose sacred mountain we are privileged to be guests.

%\clearpage
%\bibliography{../mosdef_sii}
%\bibliographystyle{apj}

\end{document}